\documentclass[cameraready]{Interspeech}

\title{Robust Multi-Tier Infant-Centered Audio Understanding with Whisper via Structured Speaker Conditioning}

\author[affiliation={1}]{Xulin}{Fan}
\author[affiliation={2}]{Jialu}{Li}
\author[affiliation={3}]{Mohammad Nur Hossain}{Khan}
\author[affiliation={1}]{Kexin}{Hu}
\renewcommand{\authorsep}{,\\}
\author[affiliation={3}]{Bashima}{Islam}
\author[affiliation={1}]{Mark}{Hasegawa-Johnson}
\author[affiliation={1}]{Nancy L.}{McElwain}

\address{
    $^1$ University of Illinois Urbana-Champaign, USA \\
    $^2$ University of Arizona, USA \\
    $^3$ Worcester Polytechnic Institute, USA
}

\email{xulinf2@illinois.edu, jialuli@arizona.edu, mkhan@wpi.edu, kexinhu2@illinois.edu, bislam@wpi.edu, jhasegaw@illinois.edu, mcelwn@illinois.edu}

\usepackage[table]{xcolor}

\usepackage{comment}
\usepackage{graphicx,hyperref}

\usepackage{amsmath}   
\usepackage{amssymb}   
\usepackage{multirow}

\begin{document}
%
\maketitle
\begin{abstract}
Recent advances in model design and self-supervised audio representations have improved speech and audio understanding, yet infant-centered naturalistic recordings remain challenging due to limited labeled data, low signal-to-noise ratio, and cross-family domain shifts. We present a family-conditioned, multi-tier audio tagger that combines a LoRA-finetuned Whisper encoder with a lightweight, target-speaker–aware Transformer for long-context inference and framewise prediction across tiers. To improve temporal coherence, we incorporate a simple sequence-level smoothing loss, and to enhance robustness across households, we introduce a factorized speaker-token design with a shared tier token and a learned family-specific offset, reducing family bias and promoting generalizable representations. Together, these choices enable efficient and effective infant-centered audio tagging of daylong audio recordings in home environments.
\end{abstract}
%
%
\section{Introduction}
\label{sec:intro}

Infant-centered audio understanding aims to characterize infant–adult interactions across naturalistic settings such as homes and clinics. Depending on the label taxonomy, the task can be formulated as speaker diarization, vocalization classification, or a combination of both. Early studies primarily relied on supervised neural models tailored to specific subtasks, including child–parent speech classification~\cite{lavechin2020childparent} and vocalization categorization~\cite{li2021vocalcls}.

Self-supervised learning (SSL) has substantially advanced speech and audio understanding by enabling models to learn robust representations from large-scale unlabeled corpora before fine-tuning on task-specific data. Foundational SSL encoders such as Wav2Vec2~\cite{baevski2020wav2vec2}, HuBERT~\cite{hsu2021hubert}, and SSAST~\cite{gong2022ssast} consistently outperform purely supervised approaches across diverse downstream tasks. These general-purpose representations have also been successfully applied to infant-centered audio tagging, where the goal is to analyze naturalistic audio recordings collected using wearable devices and identify speaker and vocalization types. Several recent works leveraging SSL frameworks for child–parent interaction analysis~\cite{li2023w2vlb, fan2025bssamba, lahiri2023sslchild} demonstrate improved robustness and performance compared with supervised baselines, highlighting the value of transferable audio representations.

An alternative paradigm bypasses additional in-domain pretraining on infant-centered home recordings by directly adapting large-scale foundational speech models~\cite{radford2023whisper, chen2022wavlm}. Trained on massive and diverse datasets, these models provide strong and noise-robust acoustic features. In particular, Whisper has been successfully adapted using parameter-efficient techniques such as low-rank adaptation (LoRA)~\cite{hu2022lora} for domain-shifted speech recognition tasks, including multilingual and child speech scenarios~\cite{song2024lorawhisper, xu2024lorawhisper2, liu2024childlorawhisper}. Beyond recognition, foundational encoders have also been explored for tagging tasks. For example, Whisper-AT~\cite{gong2023whisperat} freezes the Whisper encoder and trains lightweight fusion and classification layers, demonstrating that Whisper representations can effectively support general audio tagging with minimal task-specific overhead. Similarly, recent work investigates adapting foundation models for child–parent diarization with promising results~\cite{xu2024foundmodel}, while synthetic data augmentation has been shown to further improve performance in real-world child–parent interaction settings~\cite{xu2025dataaug}.

Despite this progress, infant-centered audio understanding remains challenging. First, unlike standard clip-level datasets such as AudioSet, naturalistic recordings require frame-level labeling at fine temporal resolution, effectively combining diarization with vocalization classification and demanding precise boundary detection. Second, recording conditions vary substantially across households, including differences in the physical environment, background noise, and speaker characteristics, which introduces significant domain shifts between families. Third, wearable microphones produce variable signal quality due to changing distances between the device and speakers, overlapping and background speech at low signal-to-noise ratios are also common, complicating reliable target-speaker analysis.

To address these challenges, we propose a multi-tier, frame-level audio tagging framework that jointly performs speaker-aware diarization and vocalization classification in naturalistic infant-centered recordings. Specifically, the model predicts, for each tier (e.g., child, female caregiver, male caregiver, sibling), a frame-aligned sequence of mutually exclusive vocalization labels while allowing simultaneous activity across tiers. Our approach integrates a LoRA-finetuned Whisper encoder with a lightweight sequence model and tier-specific classifiers, and we compare against strong SSL baselines, including Wav2Vec-based systems, to contextualize performance gains. The main contributions of this work are:


\begin{enumerate}
\item A sequence-level infant-centered audio tagging framework that combines a LoRA-adapted Whisper encoder, a projector, a target-speaker extractor, and tier-specific classifiers. Unlike prior approaches~\cite{fan2025bssamba, li2023w2vlb}, it effectively models overlapping vocalizations from multiple family members.

\item A factorized speaker-token representation consisting of a shared tier token and a family-specific offset, which is designed to separate tier-level information from family-dependent variability and empirically improves cross-family performance on our evaluation split.

\item A sequence-level auxiliary temporal smoothing loss that stabilizes predictions over time and promotes globally consistent framewise labeling.
\end{enumerate}

\begin{figure}[ht]
  \centering
  \includegraphics[width=1\linewidth]{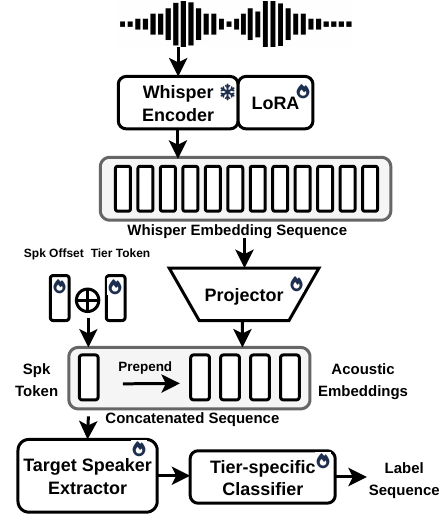}
  \vspace{-20pt}
  \caption{Training Stage of Proposed Framework}
  \label{fig:architecture}
  \vspace{-5.4pt}
\end{figure}

\section{Methods}
\label{sec:Overview}
In this section, we will outline our task formulation in section \ref{subsec:problem}, model architecture in section \ref{subsec:architecture}, and our design of the loss functions in section \ref{subsec:loss}.

\subsection{Problem Formulation}
\label{subsec:problem}
Let $\mathcal{F}$ be a set of families partitioned into training, validation, test families with no overlap. Each family $f\in\mathcal{F}$ has $R_f$ audio recordings,
indexed by $r\in\{1,\dots,R_f\}$, collected as $\mathcal{X}_f=\{\mathbf{x}^{(f,r)}\}_{r=1}^{R_f}$,
where $r$ denotes the $r$-th clip from family $f$.
Fix a frame duration $\Delta>0$ (seconds). For a recording $\mathbf{x}^{(f,r)}$ with duration
$T$ seconds, the number of frames is \(L^{(f,r)}=\bigl\lceil T/\Delta \bigr\rceil\).

We use a fixed set of tiers \(\mathcal{T}=\{\text{CHN},\text{FAN},\text{MAN},\text{CXN}\}\),
corresponding to the child, female caregiver, male caregiver, and sibling tiers, respectively.
Each tier \(\tau\in\mathcal{T}\) has its own label set \(\mathcal{V}_\tau\), where INACTIVE denotes no activity in that tier.
For example, the child tier uses
\[
\mathcal{V}_{\text{CHN}}
=\{\text{INACTIVE},\allowbreak \text{BAB},\allowbreak \text{CRY},\allowbreak \text{FUS}\},
\]
where BAB, CRY, and FUS denote babbling, crying, and fussing. Other label abbreviations used in this work include ADS, CDS, SNG, LAU, and CXN for adult-directed speech, child-directed speech, singing/rhythmic speech, laughter, and sibling vocalization.
Labels are chosen per frame and are mutually exclusive within a tier, while different tiers may be active simultaneously.

Given a clip and its known family ID \((\mathbf{x}^{(f,r)},\, f)\), the task is to output one
frame-aligned label sequence per tier $\hat{\mathbf{y}}^{(\tau)}_{1:L^{(f,r)}}$:
\[
g \colon (\mathbf{x}^{(f,r)}, f)\longmapsto
\left\{\hat{\mathbf{y}}^{(\tau)}_{1:L^{(f,r)}}\right\}_{\tau\in\mathcal{T}},
\qquad
\hat{\mathbf{y}}^{(\tau)}_{1:L^{(f,r)}} \in \mathcal{V}_\tau.
\]

\begin{table*}[t]
\centering
\setlength{\tabcolsep}{8pt}
\renewcommand{\arraystretch}{1.2}
\begin{tabular}{l|c|ccc|cccc|cc|c}
\hline
\multirow{2}{*}{Split}
& \multirow{2}{*}{\#Fam.}
& \multicolumn{3}{c|}{CHN}
& \multicolumn{4}{c|}{FAN}
& \multicolumn{2}{c|}{MAN}
& \multicolumn{1}{c}{CXN} \\
\cline{3-12}
&
& BAB & CRY & FUS
& ADS & CDS & SNG & LAU
& ADS & CDS
& CXN \\
\hline
Train & 37 & 3857.7 & 415.6 & 2280.4 & 2457.1 & 2994.7 & 743.0 & 146.6 & 1119.9 & 679.2 & 1340.2 \\
Val   & 5  & 539.9  & 83.2  & 589.8  & 418.1  & 412.2  & 133.3 & 13.0  & 281.8  & 35.6  & 164.2 \\
Test  & 10 & 1285.5 & 160.9 & 781.7  & 736.1  & 919.1  & 235.4 & 69.9  & 237.8  & 123.3 & 412.3 \\
\hline
\end{tabular}
\caption{Detailed statistics of the training, validation, and testing splits for annotated data. 
Each column reports the total annotated duration for each active vocalization labels (in seconds). 
Columns from left to right are: number of families; 
child vocalizations (CHN: BAB for babbling, CRY for crying, FUS for fussing); 
female caregiver vocalization (FAN: ADS for adult-directed speech, CDS for child-directed speech, SNG for singing/rhythmic speech, LAU for laughter); 
male caregiver vocalization (MAN: ADS for adult-directed speech, CDS for child-directed speech); 
and sibling vocalizations (CXN).}
\label{tab:dataset}
\vspace{-0.8cm}
\end{table*}

\subsection{Model Architecture}
\label{subsec:architecture}

Our model is designed to produce framewise, multi-tier tag sequences from a single audio clip. As illustrated in Fig~\ref{fig:architecture}, the architecture consists of four main components: a Whisper encoder backbone, an MLP downsampler, a target speaker extractor, and per-tier framewise classifiers.

\textbf{Whisper Encoder.}
The acoustic backbone of our model is a pre-trained Whisper-large-v2 encoder~\cite{radford2023whisper}. It processes the raw input waveform into a sequence of $D$-dimensional hidden representations, capturing rich acoustic information from the audio. The whisper encoder is finetuned with Low rank Adaptation (LoRA) during training.

\textbf{MLP Projector.}
Motivated by Slam-LLM~\cite{ma2024slamllm}, we apply a non-overlapping, windowed MLP to the encoder's output to reduce sequence length for computational efficiency and aggregate local context. Consecutive frames are grouped into windows of size $w$, flattened, and then projected by a two-layer MLP. This process yields a shorter feature sequence $Z \in \mathbb{R}^{T' \times D'}$, which is shared across all subsequent tiers.

\textbf{Target Speaker Extractor.}
This module refines the shared acoustic features $Z$ to focus on a specific target speaker for each tier $\tau$ (i.e., CHN, FAN, MAN, CXN). This is achieved by conditioning a lightweight Transformer encoder on a family-aware speaker token.

The speaker token is constructed from two learnable components. First, each tier is associated with a tier token $s_\tau$, which represents the average speaker profile for that category (e.g., the general vocal characteristics of a child). Second, to account for systematic variation across training families, such as recording conditions or speaker characteristics, we introduce a speaker offset $o_{\tau,f}$ which is unique to a specific tier in a specific family.

The final family-aware speaker token is
\[
\tilde{s}_{\tau,f} = s_\tau + o_{\tau,f}.
\]

This factorized design encourages the tier token $s_\tau$ to capture family-invariant speaker characteristics, while the offset absorbs family-dependent variability. As a result, the shared tier tokens become more robust and generalizable representations across households. This token is prepended to the downsampled sequence $Z$, forming the input $[\tilde{s}_{\tau,f};\,Z]$ for the Transformer encoder. 

After processing, the output corresponding to the prepended speaker token is discarded. The remaining outputs form a tier-specific, frame-aligned feature sequence $U_\tau \in \mathbb{R}^{T' \times D'}$. This module effectively acts as a speaker extractor, steering the acoustic representation to highlight features relevant to the intended speaker tier in a specific family.

\textbf{Per-Tier Framewise Classifiers.}
Finally, for each tier $\tau$, a dedicated two-layer MLP serves as a framewise classifier. It takes the tier-specific sequence $U_\tau$ as input and, at each time step, produces the final classification labels for that tier.

\subsection{Training Objective}
\label{subsec:loss}

Because training and test families do not overlap, speaker offsets are learned only for training families. 
At inference time, offsets are set to zero so that predictions rely solely on the shared tier tokens, which represent family-invariant speaker characteristics learned during training.

Let $Z^{(\tau)}_{t}$ denote the logits at frame $t$ for tier $\tau$, and define the posterior as
\[
\boldsymbol{\pi}^{(\tau)}_{t}
= \operatorname{softmax}\bigl(Z^{(\tau)}_{t}/T\bigr)
\]
where $T>0$ is a temperature parameter. In addition to the standard framewise cross-entropy (CE) loss for each tier, we introduce a sequence-level temporal smoothing loss:
\[
\mathcal{L}^{(\tau)}_{\mathrm{smooth}}
= \frac{1}{L-1}\sum_{t=1}^{L-1}
\lVert \boldsymbol{\pi}^{(\tau)}_{t+1}-\boldsymbol{\pi}^{(\tau)}_{t} \rVert_2^2.
\]

Empirically, we observe that framewise predictions may oscillate between labels within a single vocalization segment when the model is uncertain. The smoothing loss mitigates this behavior by penalizing rapid posterior changes across adjacent frames, thereby encouraging temporally coherent predictions that better reflect the duration structure of vocalizations.

The overall training objective is
\[
\mathcal{L}_{\text{train}}
=\frac{1}{|\mathcal{T}|}\sum_{\tau\in\mathcal{T}}
\Bigl(\mathcal{L}^{(\tau)}_{\mathrm{CE}}+\lambda\,\mathcal{L}^{(\tau)}_{\mathrm{smooth}}\Bigr).
\]

\section{Experiments}
\label{sec:experiments}

\subsection{Dataset}
\label{sec:dataset}
We conducted experiments on approximately 17 hours of labeled audio recordings from 52 families, collected using the LittleBeats\texttrademark~\cite{islam2024littlebeats} wearable device. Infants (55 percent female) were between 4 and 15 months of age (Mean = 8.03 months). The data were partitioned into training (37 families), validation (5 families), and test (10 families) sets with no family overlap. Audio recordings were annotated by trained research assistants using Praat software~\cite{boersma2011praat}. All files were double annotated, and Cohen's kappa was 0.80 or higher for all codes.

For this study, we defined four annotation tiers \(\mathcal{T}=\{\text{CHN}, \text{FAN}, \text{MAN}, \text{CXN}\}\), each with its own set of labels:
\[
\mathcal{V}_{\text{CHN}}
=\{\text{INACTIVE},\allowbreak \text{BAB},\allowbreak \text{CRY},\allowbreak \text{FUS}\},
\]
\[
\mathcal{V}_{\text{FAN}}
=\{\text{INACTIVE},\allowbreak \text{ADS},\allowbreak \text{CDS},\allowbreak \text{SNG},\allowbreak \text{LAU}\},
\]
\[
\mathcal{V}_{\text{MAN}}
=\{\text{INACTIVE},\allowbreak \text{ADS},\allowbreak \text{CDS}\},
\]
\[
\mathcal{V}_{\text{CXN}}
=\{\text{INACTIVE},\allowbreak \text{CXN}\}.
\]

The total annotated durations for each active label across the train, validation, and test sets are summarized in Table~\ref{tab:dataset}.

\begin{table*}[t]
\centering
\setlength{\tabcolsep}{3pt} 
\renewcommand{\arraystretch}{1.2}
\begin{tabular}{p{2.5cm}|cc|cc|cc|cc|cc}
\hline
\multirow{2}{*}{\textbf{Methods}} 
& \multicolumn{2}{c|}{AVG} 
& \multicolumn{2}{c|}{CHN} 
& \multicolumn{2}{c|}{FAN} 
& \multicolumn{2}{c|}{MAN} 
& \multicolumn{2}{c}{CXN} \\
\cline{2-11}
& Macro-F1 & Kappa 
& Macro-F1 & Kappa 
& Macro-F1 & Kappa 
& Macro-F1 & Kappa 
& Macro-F1 & Kappa \\
\hline
$\text{TL-TR}_{512}$~\cite{gong2023whisperat} & 69.55 & 64.04 & 58.42 & 69.01 & 69.76 & 67.91 & 72.77 & 64.61 & 77.26 & 54.63\\
$\text{W2V-LB}$~\cite{li2023w2vlb} & 67.27 & 59.27 & 68.72 & 71.97 & 61.90 & 58.79 & 61.07 & 51.48 & 77.39 & 54.81\\
\textcolor{gray}{$\text{W2V-LB*}$~\cite{li2023w2vlb}} & \textcolor{gray}{68.11} & \textcolor{gray}{60.68} & \textcolor{gray}{68.72} & \textcolor{gray}{71.97} & \textcolor{gray}{63.24} & \textcolor{gray}{60.73} & \textcolor{gray}{61.54} & \textcolor{gray}{52.11} & \textcolor{gray}{78.95} & \textcolor{gray}{57.90}\\ \hline
Proposed & \underline{74.88} & \underline{68.14} & \underline{69.13} & \underline{72.25} & \underline{71.60} & \underline{69.12} & \textbf{79.92} & \textbf{73.34} & \underline{78.87} & \underline{57.85} \\ 
w/o LoRA  & 70.45 & 63.37 & 62.64 & 70.51 & 69.82 & 67.04 & 71.72 & 60.53 & 77.66 & 55.39 \\
w/o $o_{\tau, f}$ & 73.32 & 67.20 & 66.64 & 72.37 & 71.30 & 68.86 & 77.34 & 71.47 & 78.00 & 56.11 \\
w/o $\mathcal{L}_{\mathrm{smooth}}$ & 72.66 & 66.44 & 65.72 & 71.94 & 71.00 & 69.12 & 76.61 & 69.95 & 77.31 & 54.76 \\
Proposed+TTA\cite{lin2022suta} & \textbf{74.94} & \textbf{68.24} & \textbf{69.17} & \textbf{72.33} & \textbf{71.65} & \textbf{69.15} & \underline{79.89} & \underline{73.32} & \textbf{79.03} & \textbf{58.16} \\ 
\hline
\end{tabular}
\caption{\textbf{Test performance by tier.} Macro-F1 and Cohen’s $\kappa$ (higher is better) for the proposed model and ablations. \textbf{AVG} is the unweighted mean across tiers (CHN, FAN, MAN, CXN). “w/o LoRA” disables LoRA fine-tuning of the Whisper encoder; “w/o $o_{\tau,f}$” removes family-specific offsets; “w/o $\mathcal{L}_{\mathrm{smooth}}$” removes the temporal smoothing loss. \textbf{Bold} indicates the best score in each column and \underline{underlined} values indicate the second-best. \textcolor{gray}{Gray} values correspond to results reported under an easier evaluation setting and are therefore excluded when determining the best and second-best scores. Both Macro-F1 and Kappa are reported in percentage (\%).}
\vspace{-0.8cm}
\label{tab:multi_model_results}
\end{table*}

\subsection{Experiment Configuration}
Our model takes 30-second audio samples as input. 
During training, we randomly select 30-second intervals from the recordings to increase coverage of diverse acoustic contexts. 
For evaluation and testing, the recordings are divided into non-overlapping 30-second segments.

The Whisper encoder produces frame-level embeddings of dimension $D=1280$. 
These embeddings are projected by an MLP that compresses every $w=5$ consecutive frame into a single embedding of size $D'=512$, thereby reducing the temporal resolution by a factor of five. 
The target-speaker extractor consists of a two-layer Transformer encoder with 8 attention heads, a feed-forward dimension of $4D'=2048$, a dropout rate of $0.1$, and sinusoidal positional encodings. 
Tier tokens are randomly initialized, and all speaker offsets are initialized to zero.

For training, we apply low-rank adaptation (LoRA)~\cite{hu2022lora} to the query and value projections of the Whisper encoder, using rank $r=4$ and scaling factor $\alpha=8$. 
All models are trained for 20 epochs with the Adam optimizer at a learning rate of 0.001, with the temporal smoothing loss weighted by $\lambda=0.2$. 
The checkpoint achieving the highest kappa score on the validation set is selected for final evaluation. Training requires approximately three hours on a single NVIDIA A100 GPU.

\subsection{Results}

Table~\ref{tab:multi_model_results} reports per-tier Macro-F1 and Cohen’s $\kappa$ together with across-tier averages (AVG) for our method, external baselines, and ablations. For each tier, both metrics are computed over all classes in that tier, including the INACTIVE class.

\textbf{Baselines.}
We compare against two prior approaches. First, we adapt the Time and layer-wise Transformer (TL-TR) with 512 intermediate dimension from Whisper-AT~\cite{gong2023whisperat} for framewise prediction by modifying its temporal resolution (reducing downsampling from $20\times$ to $5\times$ and removing the second downsampling stage) and attaching the same tier-wise classification heads used in our system. 

Second, we include results from W2V-LB~\cite{li2023w2vlb}, which is based on wav2vec~2.0\cite{baevski2020wav2vec2} representations pretrained on large-scale family audio. Specifically, we use the model variant that aggregates representations from all 12 transformer layers using a learned weighted average and is pretrained on approximately 4300 hours of unlabeled home recordings. Because W2V-LB is designed for single-tier prediction, it cannot model simultaneous active tiers. Therefore, we evaluate it under two protocols. In both cases, training data are preprocessed to remove overlapping active tiers using a fixed priority order (CHN $>$ FAN $>$ MAN $>$ CXN). For evaluation, \textsc{W2V-LB*} reports performance on test labels processed with the same overlap-removal rule, reflecting its intended operating condition. In contrast, \textsc{W2V-LB} is evaluated on the original multi-tier test labels, providing a direct comparison with our method but necessarily penalizing predictions when multiple tiers are active simultaneously.

\textbf{Overall comparison.}
Our proposed method achieves the best overall performance, reaching 74.88 Macro-F1 and 68.14 $\kappa$ averaged across tiers. Under the original multi-tier evaluation, it outperforms TL-TR and W2V-LB on the across-tier average and on each tier. The consistent gains indicate that the architecture effectively captures both acoustic structure and speaker-specific characteristics while remaining robust to household variability.

\textbf{Tier-wise analysis.}
The largest performance gaps between our method and W2V-LB occur in the adult tiers (FAN and MAN), whereas the difference on the child tier (CHN) is comparatively small under the same multi-tier evaluation setting. One possible explanation is pretraining bias: Whisper is trained on massive speech corpora dominated by adult speech, which may provide stronger representations for adult vocalizations, while its representations for infant and child speech from a close-mic remain relatively out-of-domain. In contrast, W2V-LB is pretrained on large-scale family recordings containing child vocalizations, which may explain its competitive performance on CHN despite its architectural limitations. This observation highlights the importance of both pretraining data distribution and model architecture when addressing domain-specific speech understanding tasks.

\textbf{Effect of architectural components.}
The ablation results clarify the role of each component. Removing LoRA (w/o LoRA) leads to consistent degradation across tiers, confirming that parameter-efficient fine-tuning of the Whisper encoder substantially improves in-domain acoustic representations. Eliminating the family offset (w/o $o_{\tau,f}$) primarily affects tiers with higher cross-family variability, especially MAN, suggesting that the offset absorbs family-dependent variability during training and allows the shared tier tokens to learn more invariant representations. Removing temporal smoothing (w/o $\mathcal{L}_{\mathrm{smooth}}$) reduces temporal stability, with larger drops in CHN and MAN tiers. Overall, these findings indicate that performance gains arise not only from stronger pretrained features but also from architectural inductive biases tailored to multi-speaker naturalistic recordings, where separating shared representations from nuisance variability and enforcing temporal consistency are both critical for robust infant-centered audio understanding.

\section{Test-Time Adaptation of the offset}

Although the proposed framework demonstrates strong performance across families, it does not explicitly adapt the family-specific offset parameters for unseen test households. In the current design, offsets are learned only for training families, and inference on new families relies solely on the shared tier tokens. While this strategy promotes generalization and avoids additional computation at test time, it may limit the model’s ability to fully capture household-specific acoustic characteristics such as recording conditions, speaker traits, or environment-dependent noise patterns of the test families.

We conducted preliminary experiments exploring unsupervised test-time adaptation (Proposed+TTA in Table~\ref{tab:multi_model_results}) using a SUTA-style objective~\cite{lin2022suta} to update offsets for unseen families.
At test time, we adapt only the speaker-offset table $\{O_{\tau,f}\}$ for unseen families, while freezing all other parameters. 
Following SUTA~\cite{lin2022suta}, we employ a tier-wise weighted combination of entropy minimization (EM) and minimum class confusion (MCC) objectives for calibration:
\[
\mathcal{L}_{\mathrm{cal}}
=\sum_{\tau\in\mathcal{T}}
\Bigl(\eta\,\mathcal{L}^{(\tau)}_{\mathrm{EM}}+(1-\eta)\,\mathcal{L}^{(\tau)}_{\mathrm{MCC}}\Bigr),\quad \eta\in[0,1].
\]
As in the original work, EM excludes frames dominated by the CTC blank class in ASR. 
Analogously in our setting, $\mathcal{L}^{(\tau)}_{\mathrm{EM}}$ is computed only over frames whose most probable prediction corresponds to an active vocalization class. We apply the test time adaptation for 5 epochs with a mixing weight $\eta=0.3$. However, the observed improvements were modest. These results suggest that, although adaptation has potential, more effective or stable approaches are needed to make it practically beneficial.

Developing robust unsupervised adaptation mechanisms for unseen families remains an important direction for future work. In particular, methods that can reliably personalize representations without supervision, while preserving stability and efficiency, could further improve performance in realistic deployment scenarios where domain shifts are inevitable.

\section{Conclusion}
\label{sec:conclusion}
We presented a compact framework for infant-centered multi-tier audio tagging that unifies diarization and vocalization classification. By combining a LoRA-finetuned Whisper encoder with structured speaker conditioning and tier-specific heads, the model supports overlapping speakers and framewise prediction. Experiments on a family-disjoint split show consistent empirical gains over baselines, suggesting the effectiveness of pretrained representations with task-specific inductive biases for naturalistic infant-centered recordings.

\section{Generative AI Use Disclosure}

We used Claude and GPT for language editing and manuscript polishing, including improving clarity of expression, correcting grammatical errors, and formatting \LaTeX{} tables. All technical content, experimental design, analysis, and scientific contributions are entirely the work of the authors. The authors have carefully reviewed the manuscript and take full responsibility for its content.
\section{Acknowledgement}
This study was supported by funding from the National Institute on Drug Abuse (R34DA050256; R01DA059422). For the experiments presented here, we used the Delta System at the National Center for Supercomputing Applications through ACCESS allocations CIS240417 and CIS250040.

\bibliographystyle{IEEEtran}
\bibliography{mybib}

\end{document}